\RequirePackage[2024-06-01]{latexrelease}
\documentclass[aip,rsi,amsmath,amssymb,preprint]{revtex4-2}

\usepackage{etoolbox}
\AtBeginEnvironment{table}{\linespread{1}\selectfont}   
\usepackage[T1]{fontenc}
\usepackage{graphicx}
\graphicspath{{figures/}}
\usepackage{booktabs}
\usepackage{tabularx}
\usepackage{eurosym}
\usepackage{textcomp}
\usepackage{float}
\usepackage{placeins}
\usepackage{url}

\begin{document}

\title{Low-Cost Spectrophotometers: A Comparative Evaluation of Open-Source Architectures}

\author{Alex M. Rivera-Rivera}
\affiliation{Innovation Laboratory, Technological University of Havana Jos\'e Antonio Echeverr\'ia (CUJAE), 19390 Havana, Cuba}
\affiliation{Facult\'e des Sciences, Universit\'e Libre de Bruxelles (ULB), 1050 Brussels, Belgium}

\author{Daniela Martinez-Ortiz}
\affiliation{Center for Complex Systems, Physics Faculty, University of Havana, 10400 Havana, Cuba}
\affiliation{Facult\'e des Sciences, Universit\'e Libre de Bruxelles (ULB), 1050 Brussels, Belgium}

\author{Fran\c{c}ois Quitin}
\affiliation{Brussels School of Engineering, Universit\'e libre de Bruxelles, Avenue Franklin Roosevelt 50, 1000 Brussels, Belgium}

\author{Denis Terwagne}
\affiliation{Frugal Lab, D\'epartement de physique, Facult\'e des Sciences, Universit\'e libre de Bruxelles (ULB), 1050 Brussels, Belgium}
\affiliation{Fab Lab ULB, Universit\'e libre de Bruxelles (ULB), 1050 Brussels, Belgium}

\author{Stamatios C. Nicolis}
\affiliation{Unit of Biological and Artificial Self-organizing Systems, Department of Physical Chemistry and Theoretical Biology, Center for Nonlinear Phenomena and Complex Systems, Universit\'e libre de Bruxelles, Bd du Triomphe, 1050 Brussels, Belgium}

\author{Ernesto Altshuler}
\email[Author to whom correspondence should be addressed: ]{ealtshuler@fisica.uh.cu}
\affiliation{Center for Complex Systems, Physics Faculty, University of Havana, 10400 Havana, Cuba}

\author{Alexandre Campo}
\email[Author to whom correspondence should be addressed: ]{alexandre.campo@ulb.be}
\affiliation{Unit of Biological and Artificial Self-organizing Systems, Department of Physical Chemistry and Theoretical Biology, Center for Nonlinear Phenomena and Complex Systems, Universit\'e libre de Bruxelles, Bd du Triomphe, 1050 Brussels, Belgium}

\begin{abstract}
The rapid growth of low-cost and open-source spectrophotometers has created a need for systematic methodologies to evaluate and compare alternative instrument architectures. This study presents a comparative assessment of three representative open-source spectrophotometric platforms, encompassing discrete multispectral sensing and continuous-spectrum imaging approaches. Their performance was evaluated against a commercial Thermo Scientific Genesys 30 spectrophotometer in terms of spectral agreement, portability, implementation complexity, and cost. The results demonstrate that no single architecture is universally optimal; instead, each presents distinct trade-offs that determine its suitability for specific applications. Multispectral sensor-based systems offer compact, low-power solutions for field deployment, whereas imaging-based platforms provide spectral performance closer to laboratory-grade instrumentation. All hardware designs, firmware, and software developed in this work are released as open-source resources to support reproducibility and further development. The study highlights the importance of matching instrument architecture to application requirements and proposes a practical framework for the evaluation and selection of low-cost spectrophotometers, facilitating the adoption of open-source analytical tools for water quality monitoring and related applications.
\end{abstract}

\maketitle

\section{Introduction}\label{sec1}

Spectrophotometry is one of the most widely used analytical techniques for the qualitative and quantitative characterization of chemical compounds, with applications ranging from environmental monitoring and water quality assessment to biomedical diagnostics, agriculture, food analysis and the control of industrial processes \citep{yoon2016introduction,yu2023open}. By measuring the interaction between light and matter, spectrophotometric methods provide rapid, non-destructive, and highly versatile analytical information, making them essential tools in both research and routine laboratory practice \citep{malvandi2022application,workman2026advances}.

Despite its widespread adoption, conventional laboratory spectrophotometers remain relatively expensive, bulky, and dependent on stable power infrastructure, which limits their deployment in field environments, educational laboratories, and resource-constrained settings \citep{gonzalez2020development,pereira2019low,shidiq2020simple}. These limitations have driven growing interest in the development of low-cost and open-source alternatives that aim to preserve acceptable analytical performance while significantly reducing acquisition and operational costs \citep{laganovska2020portable,wenzel2023open}. Recent advances in additive manufacturing, open-source electronics, low-cost optical components, and single-board computing platforms have further accelerated this trend, enabling the development of portable and customizable spectrophotometric systems accessible to researchers, educators, and citizen-science communities \citep{feng2021open,lopez2023design}.

The development of low-cost spectrophotometers can be understood in terms of modular subsystems, including light sources, dispersive or filtering elements, detectors, sample holders, and processing units. The selection of each component significantly influences the cost, spectral performance, complexity, and reproducibility of the system. Table~\ref{tab:components} summarises the main options reported in the literature, together with their respective advantages and limitations.

\begin{table}[!htbp]
\caption{Components commonly used in low-cost spectrophotometers}\label{tab:components}
\centering
\footnotesize
\setlength{\tabcolsep}{4pt}
\begin{tabularx}{\textwidth}{@{}>{\raggedright\arraybackslash\hspace{0pt}}p{0.135\textwidth} >{\raggedright\arraybackslash\hspace{0pt}}X >{\raggedright\arraybackslash\hspace{0pt}}X >{\raggedright\arraybackslash\hspace{0pt}}X@{}}
\toprule
Subsystem & Option & Advantages & Limitations \\
\midrule
Light source & LEDs (white, RGB, UV or specific colours) & Low cost, widely available \citep{granica2019analytical} & Intensity varies; white LEDs cover the visible range unevenly \citep{granica2019analytical} \\
 & Xenon or tungsten lamps & Broad continuous spectrum, covers the UV-Vis range evenly \citep{visconti2017hardware} & Costly and bulky; high-voltage supply \citep{visconti2017hardware} \\
\midrule
Dispersing element & Diffraction grating & Low cost, easy to obtain \citep{granica2019analytical} & Quality and efficiency inconsistent \citep{takala2022reproducibility} \\
 & Monochromator (rotating prism) & High resolution; selects one wavelength at a time \citep{takala2022reproducibility} & Costly and bulky; high-voltage supply \citep{visconti2017hardware} \\
 & Optical filters (band-pass, long-pass) & Inexpensive and easy to implement; effective in fluorometry \citep{jeong2021open} & Fixed band per filter; not flexible \citep{jeong2021open} \\
\midrule
Sample holder & Standard cuvettes (plastic, glass, quartz) & Laboratory standard; the 10~mm path simplifies Beer-Lambert calculations \citep{pitula2025glass} & Large sample volume; bubbles form easily; quartz is expensive \citep{pitula2025glass} \\
 & Capillaries or hollow-core fibres (ARHCF) & Microlitre sample volumes \citep{pitula2025glass} & Capillary action weakens below 5~\textmu L; complex setup \citep{pitula2025glass} \\
\midrule
Detector & Smartphone camera or webcam (CMOS, CCD) & Very low cost, widely available \citep{granica2019analytical} & Bayer filter causes spectral overlap; automatic processing adds error \citep{granica2019analytical} \\
 & Dedicated spectral sensors (C12880MA, AS7341, AS7265x) & High sensitivity, stability and reproducibility \citep{laganovska2020portable} & Higher cost; fixed resolution, e.g. 15~nm \citep{laganovska2020portable} \\
 & Photoresistor or photodiode & Very low cost and very simple \citep{yu2023open} & No wavelength information, only total intensity \citep{yu2023open} \\
\midrule
Housing & 3D printing (PLA, ABS) & Rapid prototyping, custom geometries \citep{wang20193d} & Print precision affects performance; rough surfaces scatter \citep{wang20193d} \\
 & Laser cutting (acrylic, MDF) & Precise assembly \citep{feng2021open} & Needs a laser cutter; limited to 2.5D shapes \citep{feng2021open} \\
\midrule
Control and processing & Microcontroller (Arduino, Teensy, ESP32) & Low cost, open source, large community \citep{granica2019analytical} & Less processing power than an SBC or a PC \citep{takala2022reproducibility} \\
 & Single-board computer (Raspberry Pi) & Runs image analysis and a graphical interface on board \citep{takala2022reproducibility} & Higher cost and power consumption \citep{takala2022reproducibility} \\
 & Smartphone & No additional control hardware \citep{granica2019analytical} & Depends on the phone model; limited low-level camera control \citep{yu2023open} \\
\bottomrule
\end{tabularx}
\end{table}

The first is imaging-based spectroscopy, where a diffraction grating disperses light into its spectral components, which are then captured by a camera sensor and processed computationally to reconstruct a continuous spectrum. These systems can achieve relatively high spectral resolution at low cost and benefit from the processing capabilities of smartphones or single-board computers \citep{das2016ultra}.

The second approach relies on integrated multispectral sensors, such as the AS7341 or AS7265x families, which incorporate optical filters directly on the detector. These devices measure discrete wavelength bands without requiring complex optical assemblies, offering advantages in portability, robustness, and ease of calibration, although typically at the expense of spectral resolution \citep{daurai2023comparison,yu2023open}.

Numerous prototypes based on both architectures have been reported and validated against commercial spectrophotometers, demonstrating their feasibility for analytical applications \citep{feng2021open,gonzalez2020development,laganovska2020portable,yu2023open}. However, most studies focus on individual implementations or specific use cases, which makes direct comparison between architectures difficult. As a result, objective criteria for selecting the most appropriate design for a given application remain underdeveloped.

Representative implementations of DIY spectrophotometers further illustrate this diversity in design choices, including variations in optical configuration, detection strategy, and computational processing; Table~\ref{tab:diy} provides an overview of representative systems reported in the literature.

\begin{table}[!htbp]
\caption{Representative open-source and DIY spectrophotometer projects reported in the literature}\label{tab:diy}
\centering
\footnotesize
\setlength{\tabcolsep}{3pt}
\begin{tabularx}{\textwidth}{@{}>{\raggedright\arraybackslash\hspace{0pt}}p{0.185\textwidth} >{\raggedright\arraybackslash\hspace{0pt}}p{0.125\textwidth} >{\raggedright\arraybackslash\hspace{0pt}}p{0.125\textwidth} >{\raggedright\arraybackslash\hspace{0pt}}X >{\raggedright\arraybackslash\hspace{0pt}}X >{\raggedright\arraybackslash\hspace{0pt}}p{0.07\textwidth}@{}}
\toprule
Project & Dispersing element & Detector & Control and processing & Notes & Cost\,\textsuperscript{a)} \\
\midrule
EOSpec \citep{feng2021open} & Diffraction grating & Smartphone camera & Smartphone app: image analysis, spectrum extraction, absorbance & Double beam; sample and reference in one image; robust to ambient light & 30 USD \\
OSMS \citep{laganovska2020portable} & Reflective grating inside the chip & Hamamatsu C12880MA, 288-pixel CMOS & Arduino Nano with Bluetooth; mobile app or PC & Single beam; high measurement quality; 450--750~nm & 225 EUR \\
HSFS \citep{yu2023open} & None; filters on the sensor & AS7341, 8 channels & Arduino Nano with Bluetooth; Android app & Colorimeter and fluorimeter; portable and wireless & 50 USD \\
Portable fluorimeter \citep{lopez2023design} & Concave grating in a MEMS chip & Hamamatsu C12880MA & Arduino UNO, Mega or Zero; data to spreadsheet & High sensitivity; 405~nm laser and 365~nm UV LED & 500 EUR \\
Phone spectrophotometer \citep{pereira2019low} & Interchangeable LEDs (587, 465~nm) & Ambient light sensor of the phone & Smartphone app & Colorimeter at discrete wavelengths; extremely simple & 5 EUR \\
Arduino spectrophotometer \citep{chung_arduino_spectrophotometer} & RGB LED as selectable source & Light-dependent resistor & Arduino UNO; data on the serial monitor & Basic educational colorimeter & n.\,s. \\
PySpectrometer2 \citep{wright2022pyspectrometer2} & Commercial pocket spectroscope & Raspberry Pi or USB camera & Raspberry Pi or Linux PC & OpenCV; multi-point calibration and polynomial fit & n.\,s. \\
\bottomrule
\end{tabularx}

\vspace{2pt}
\begin{minipage}{\linewidth}\raggedright\footnotesize
\textsuperscript{a)}Approximate cost as reported in each source, in the currency given there (USD or EUR); n.\,s., not stated.
\end{minipage}
\end{table}
 Consequently, there is currently a lack of systematic evaluations that compare imaging-based and multispectral-sensor-based spectrophotometers under equivalent experimental conditions. Such comparisons are essential to understand the trade-offs between spectral resolution, precision, calibration complexity, portability, and cost.

To address this gap, this work presents a comparative evaluation of three representative low-cost spectrophotometric platforms, covering both continuous-spectrum (imaging-based) and discrete-band (multispectral) architectures. The evaluated systems comprise a custom-developed multispectral spectrophotometer based on the SparkFun Triad AS7265x sensor (System A), an imaging spectrophotometer implemented using the open-source PySpectrometer2 platform (System B), and the commercial open-source Little Garden spectrophotometer (System C). All three platforms were implemented, calibrated, and experimentally validated against a commercial Thermo Scientific Genesys 30 spectrophotometer. In addition to analytical performance, the comparison considers spectral resolution, precision, calibration requirements, portability, implementation complexity, and cost. Rather than identifying a universally superior design, the objective is to establish a practical framework for selecting low-cost spectrophotometer architectures according to the requirements of specific applications, particularly in the context of water quality monitoring.

\FloatBarrier
\section{Materials and Methods}\label{sec2}

Here, we evaluate three representative open-source spectrophotometric platforms encompassing the two principal architectural approaches identified in the literature: discrete multispectral sensing and continuous-spectrum imaging spectroscopy. The comparison considered both quantitative and qualitative factors, including spectral measurement performance, agreement with a commercial reference instrument, portability, implementation complexity, calibration requirements, scalability, and overall cost. These criteria represent key trade-offs commonly encountered in the development of low-cost spectrophotometric instrumentation for environmental monitoring, educational applications, and citizen science initiatives.

A Thermo Scientific Genesys 30 spectrophotometer (Thermo Fisher Scientific, Madison, WI, USA; wavelength range 325–1100 nm) was used as reference instrument throughout this study. The Genesys 30 is a laboratory-grade spectrophotometer widely used for absorbance measurements and analytical applications. Its established performance and reliability make it suitable as a benchmark for validating low-cost spectrophotometric platforms. All comparative measurements were performed using identical samples and measurement procedures whenever possible.

\subsection{Evaluated Spectrophotometric Platforms}\label{subsec2.1}

\begin{figure*}[!htbp]
\centering
\includegraphics[width=\linewidth, draft=false]{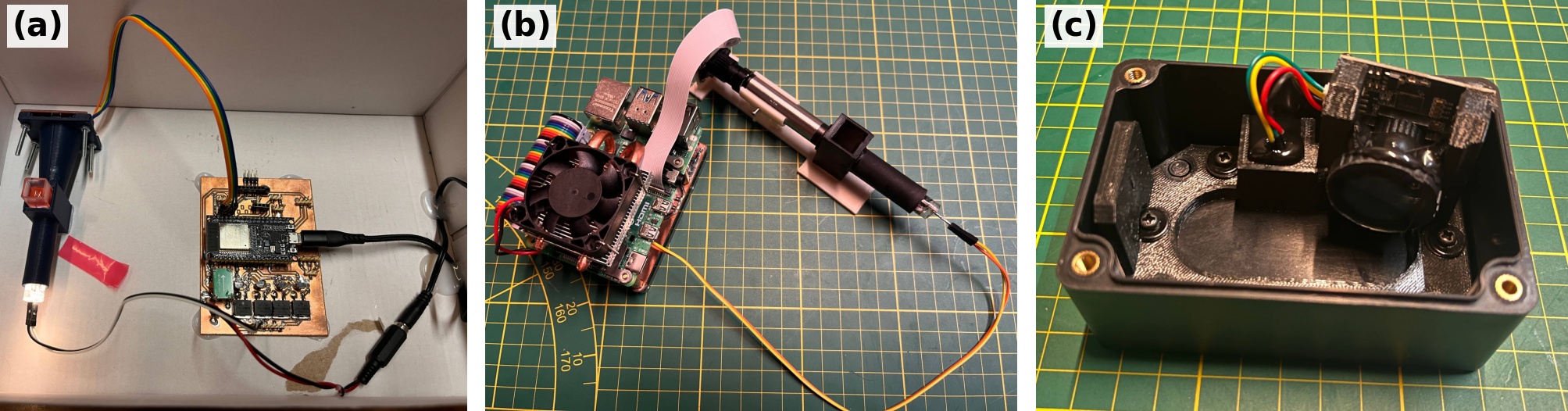}
\caption{Spectroscopic systems evaluated in this study. (a) AS7265x-based multispectral system mounted on a custom 3D-printed structure and controlled by an ESP32 microcontroller. (b) Raspberry Pi camera coupled to a commercial pocket spectroscope (Paton Hawksley Pocket Spectroscope), mounted on a custom 3D-printed structure and controlled by a Raspberry Pi 4B. (c) Little Garden spectrometer, consisting of a modified webcam with the Bayer filter removed and a homemade diffraction grating fabricated from a DVD, connected directly to a computer via USB.}
\label{fig:systems}
\end{figure*}

Figure~\ref{fig:systems} shows images of the spectrophotometric platforms used in this study.

System A [Fig.~\ref{fig:systems}(a)] was developed around the SparkFun Triad AS7265x multispectral sensor, which integrates three CMOS devices (AS72651, AS72652, and AS72653) and provides 18 discrete spectral channels distributed between 410 and 940 nm. Each channel has an approximate full-width half-height (FWHM) bandwidth of 20 nm and is defined by interferometric filters integrated directly onto the sensor, eliminating the need for external dispersive elements. The built-in illumination LEDs were disabled, and the system operated exclusively with an external light source. Data acquisition and processing were performed using an ESP32 microcontroller, which managed communication with the sensors, applied white balance correction, and calculated absorbance from the logarithmic ratio of transmitted and reference intensities. The software, developed in Python and released as open source, averages ten consecutive measurements to reduce stochastic noise and implements automatic gain control that dynamically adjusts the integration time.

System B [Fig.~\ref{fig:systems}(b)] employed an image spectroscopy approach based on a Raspberry Pi camera coupled to a commercial pocket spectroscope (Paton Hawksley Pocket Spectroscope) equipped with a transmission diffraction grating. A 12 mm M12 lens was incorporated to optimize spectral focusing on the sensor. Spectral acquisition and processing were performed using the PySpectrometer2 software, which implements spectral extraction via vertical averaging and allows multipoint wavelength calibration.

System C [Fig.~\ref{fig:systems}(c)] is the Little Garden spectrophotometer. The device uses a modified webcam, from which the Bayer filter was removed to maximize spectral sensitivity, along with a diffraction grating fabricated from a fragment of a DVD. The optical components are housed in a 3D-printed enclosure and connect directly to a computer via USB. Spectral acquisition was performed using a modified version of the Theremino Spectrometer set to a resolution of 1920 × 1080 pixels. Before each measurement, the vertical region of interest was manually adjusted to maximize spectral capture and minimize background contribution.

All systems used a 10~W halogen lamp as a light source, making them more comparable to the Genesys 30 commercial spectrophotometer.

\subsection{Calibration}\label{subsec2.2}

\begin{figure}[!b]
\centering
\includegraphics[width=0.44\linewidth, draft=false]{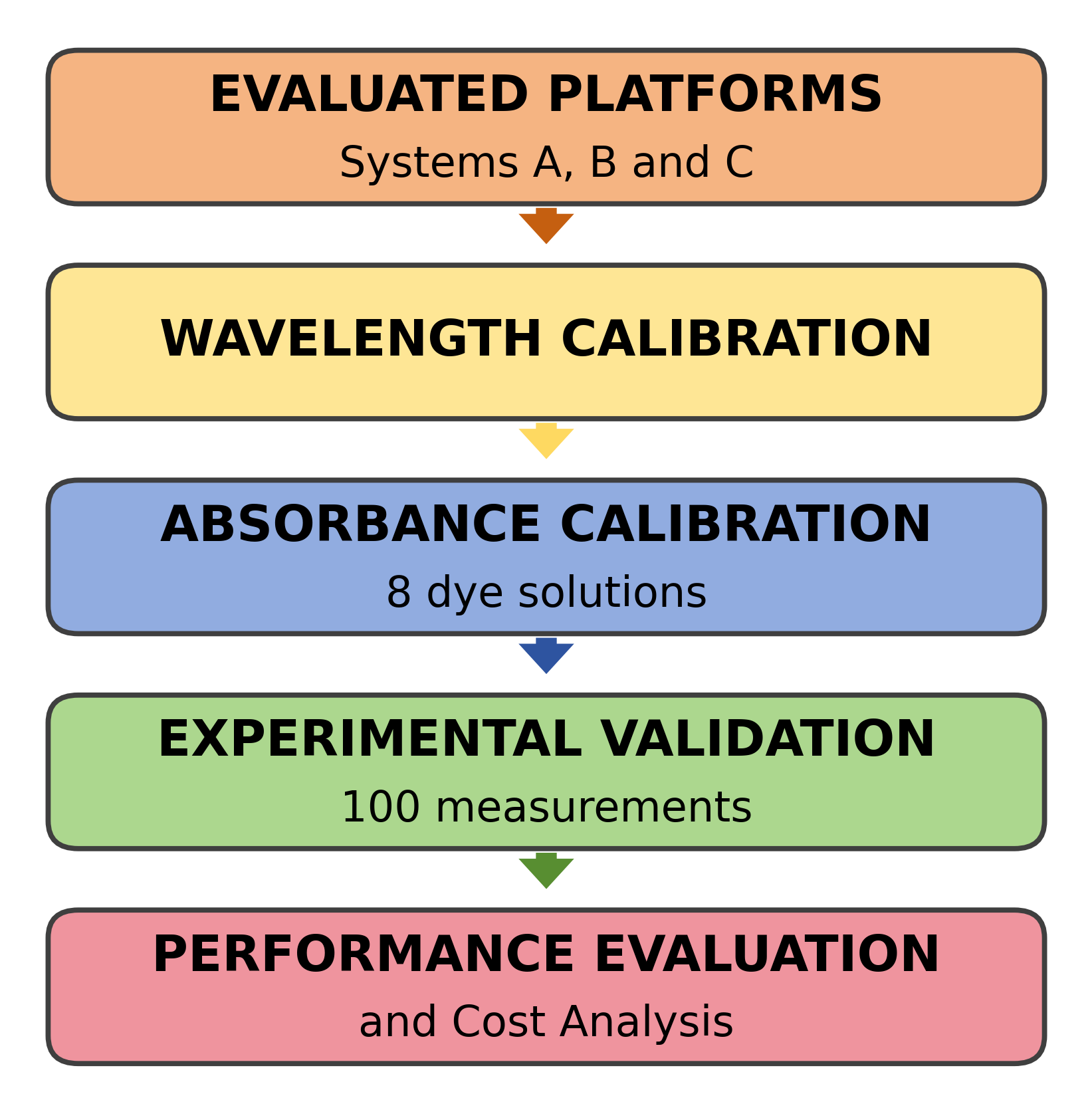}
\caption{Workflow of the wavelength calibration, absorbance calibration, and experimental validation procedures applied to the three evaluated spectrophotometric platforms.}
\label{fig:workflow}
\end{figure}

Two independent calibrations were applied to each platform: a wavelength calibration, which assigns a wavelength to every channel or pixel, and an absorbance calibration, which brings the absorbance scale of each platform onto that of the reference instrument. Both procedures, together with the experimental validation described in Section~\ref{subsec2.3}, are summarised in Fig.~\ref{fig:workflow}.

\subsubsection{Preparation of the dye solutions}\label{subsubsec2.2.1}

\begin{figure*}[!htbp]
\centering
\includegraphics[width=\linewidth, draft=false]{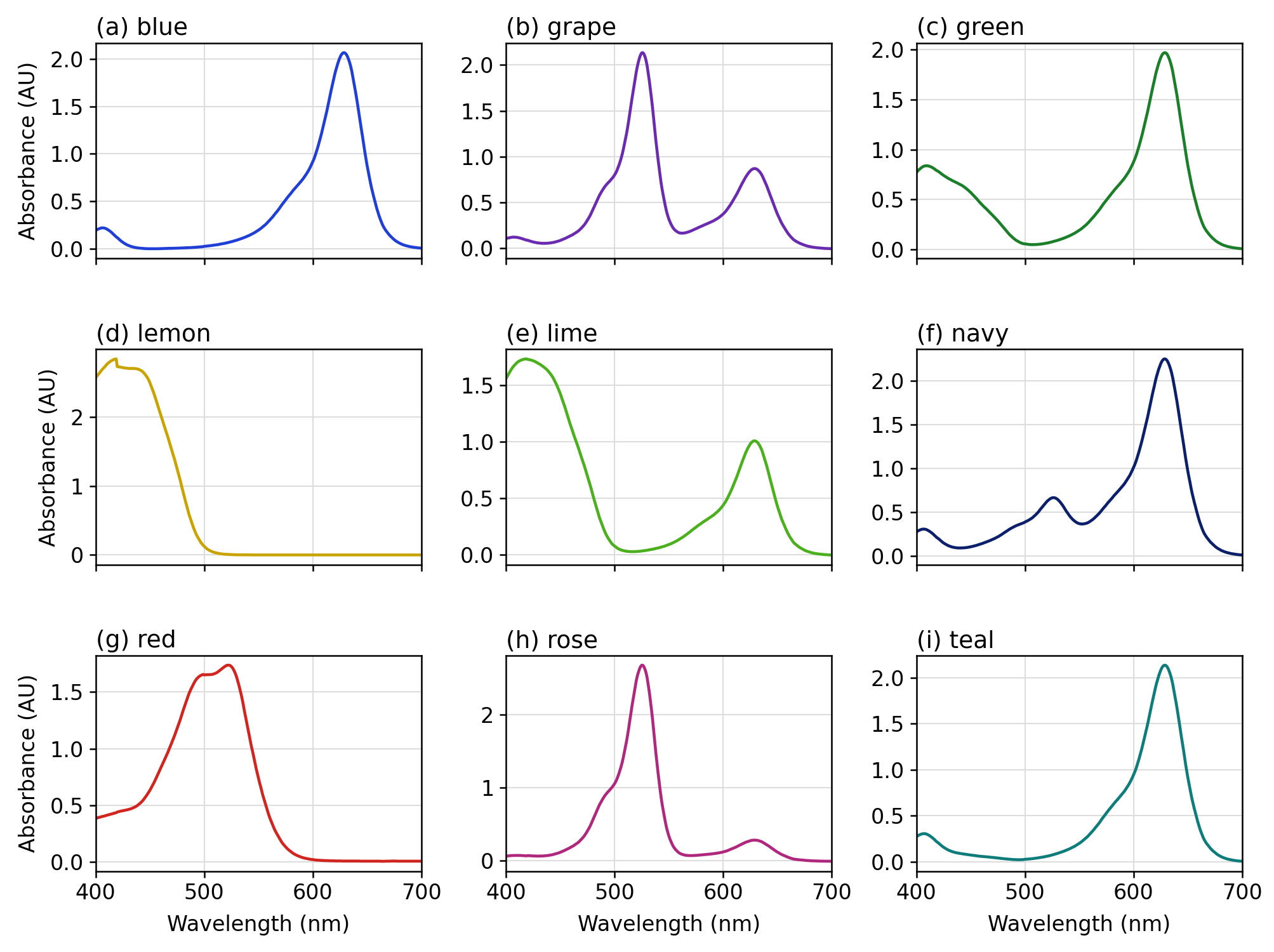}
\caption{Reference absorbance spectra, measured with the Thermo Scientific Genesys 30 spectrophotometer, of nine of the eleven dye solutions used in this study: (a) blue, (b) grape, (c) green, (d) lemon, (e) lime, (f) navy, (g) red, (h) rose and (i) teal (the orange and pink solutions are omitted).}
\label{fig:dyes}
\end{figure*}

Eleven aqueous solutions prepared from commercially available water-soluble food dyes were used. The dyes were taken from a commercial twelve-colour concentrated liquid food-colouring set (6~mL per bottle; Amazon ASIN B08X6FMHKP). The manufacturer does not disclose the identity or the concentration of the individual colorants; the dyes are therefore identified here by their commercial colour names, and their reference absorbance spectra are reported so that they can be recognised independently of the supplier. Each solution was prepared by diluting one drop of food dye in 10~mL of distilled water, which provided a representative range of absorbance levels across the visible spectrum. Eight solutions (grape, lemon, lime, navy, orange, pink, rose and teal) were used to build the absorbance calibration, and the three remaining ones (blue, green and red) were kept aside as an independent validation set. The reference absorbance spectra of nine of these solutions, measured with the Thermo Scientific Genesys 30 spectrophotometer, are shown in Fig.~\ref{fig:dyes}.

\subsubsection{Wavelength calibration}\label{subsubsec2.2.2}

System A does not require a wavelength calibration: its eighteen channels are defined by interferometric filters deposited directly on the sensor, and their nominal centre wavelengths are fixed by the manufacturer. For the two imaging platforms, the pixel-to-wavelength mapping was established from the emission lines of a mercury lamp. For System B, a third-order polynomial was fitted to four lines (404.7, 435.8, 546.1 and 577.0~nm), which compensates for the nonlinear dispersion of the grating and for the aberrations introduced by the lens. For System C, the Theremino software uses only two lines (435.8 and 546.1~nm), which fix a linear pixel-to-wavelength mapping.

\subsubsection{Absorbance calibration}\label{subsubsec2.2.3}

The three platforms were compared by measuring the same dye solutions, in the same cuvettes, as the reference spectrophotometer. Under those conditions the absorbance a platform reports and the absorbance measured by the reference instrument differ only by a gain and an offset, so the two were related by a polynomial of first order in the signal,
\begin{equation}\label{eq:cal}
A_{\text{ref}} \;=\; a_1(\lambda)\, s \;+\; a_0(\lambda),
\end{equation}
where $s$ is the absorbance reported by the platform, $a_1(\lambda)$ is the gain and $a_0(\lambda)$ the offset of that platform at the wavelength $\lambda$. Each wavelength provides one calibration point per dye, so the two parameters of the first-order polynomial are fitted to seven or eight points. A second-order polynomial in $s$ was also tested. In leave-one-dye-out cross-validation over the eight calibration dyes it did not reduce the prediction error for any platform and was unstable for System~B, so the first order was retained.

A single gain and a single offset shared by all dyes and all wavelengths do not reproduce the reference, because the deviation of each platform with respect to it depends on the wavelength. The two coefficients of Eq.~\eqref{eq:cal} were therefore treated as functions of wavelength. Rather than fitting them independently at each wavelength, which ignores the fact that neighbouring wavelengths share the same detector, the same source and the same dye, $a_1(\lambda)$ and $a_0(\lambda)$ were represented as smooth functions of wavelength and obtained in a single least-squares fit over all dyes and all wavelengths at once. The two functions were expanded in an orthogonal polynomial basis in the normalised wavelength. The degree of this expansion in $\lambda$, which is distinct from the first order in $s$, was selected on validation performance: 18 for System~B and 8 for System~C.

System A is treated separately. Its eleven channels within 400--700~nm are discrete filters rather than a sampled continuum, so no smoothing in $\lambda$ is meaningful and its gain and offset were fitted at each channel independently.

The calibration grid comprises 11 wavelengths for System A, 301 for System B and 570 for System C. The resulting calibration is nevertheless compact: 22 coefficients for System A (two for each of its 11 channels), 38 for System B and 18 for System C (two for each of the 19 and 9 terms of the expansion). For the two continuous platforms the calibration is a smooth surface $A_{\text{ref}}=f(\lambda,s)$ rather than a set of unrelated fits, so it can be evaluated at any wavelength within the calibrated range and its coefficients vary continuously, as a physically meaningful instrument response must.

Because the calibration is built from the eight calibration dyes only, it is evaluated on dyes that were not used to construct it: the blue, green and red solutions were held out for this purpose, and every value reported in Section~\ref{sec4} refers exclusively to that independent validation set.

\FloatBarrier
\subsection{Experimental Validation}\label{subsec2.3}

All measurements were performed under identical experimental conditions using standard 10 mm optical plastic cuvettes, in complete darkness, and at a controlled ambient temperature of 22 °C. White and dark references were acquired before each series of measurements.

Following the calibration procedure described in Section 2.2, the performance of the three platforms was experimentally validated on the three solutions held out for that purpose in Section~\ref{subsubsec2.2.1}: blue, green and red, whose reference spectra are among those of Fig.~\ref{fig:dyes}. For each sample and platform, 100 consecutive measurements were acquired, manually repositioning the cuvette between readings to simulate the typical operational variability of non-automated environments. These three solutions were used both for the agreement values reported below and for the qualitative assessment of the reconstructed spectral profiles.

The agreement between each low-cost platform and the Thermo Scientific Genesys 30 reference spectrophotometer was quantified using the coefficient of determination ($R^{2}$) and the adjusted coefficient of determination ($R^{2}_{adj}$), the latter computed as $R^{2}_{adj}=1-(1-R^{2})(n-1)/(n-p-1)$ with $p=1$, the order of the calibration polynomial in Eq.~\eqref{eq:cal}. Reporting the adjusted value matters here because the number of comparison points $n$ differs by more than one order of magnitude between the discrete and the continuous platforms. The total number of coefficients actually fitted per platform is reported separately in Section~\ref{subsubsec2.2.3}. For System A, the absorbance values measured at its discrete spectral channels were compared with the absorbance of the reference instrument at the same wavelengths; for Systems B and C, the reconstructed continuous spectra were compared with the reference spectra over the common spectral range. All comparisons were restricted to the 400--700~nm interval common to the three platforms.

The performance of the three platforms was evaluated using various complementary metrics. Spectral resolution was determined from the effective spectral characteristics of each system. The repeatability and stability of the measurements were assessed using the standard deviation and the coefficient of variation (CV) calculated from the 100 repetitions performed for each sample. Accuracy was quantified using the coefficient of determination (R²) between the absorbance spectra obtained by each low-cost platform and those measured with the Genesys 30 within the 400--700~nm range common to the three platforms. Finally, a qualitative evaluation of ease of calibration and maintenance was performed, considering the number of steps required, the need for external equipment, and the mechanical robustness of the optical assembly.

\FloatBarrier
\section{Results and Discussion}\label{sec4}

\begin{table}[!htbp]
\caption{Agreement with the reference spectrophotometer and repeatability of the three platforms, evaluated over the 400--700~nm range on the three validation dyes}\label{tab:metrics}
\centering
\footnotesize
\begin{tabular}{lrrrrrr}
\toprule
Platform & $R^{2}$ & $R^{2}_{adj}$\,\textsuperscript{a)} & $n$\,\textsuperscript{b)} & $N_{c}$\,\textsuperscript{c)} & SD (AU)\,\textsuperscript{d)} & CV (\%)\,\textsuperscript{e)} \\
\midrule
A (AS7265x)         & 0.805 & 0.783 & 11  & 22 & 0.055 & 8.2 \\
B (Raspberry Pi)    & 0.983 & 0.983 & 301 & 38 & 0.029 & 4.0 \\
C (Little Garden)   & 0.975 & 0.975 & 570 & 18 & 0.001 & 0.3 \\
\bottomrule
\end{tabular}

\vspace{2pt}
\begin{minipage}{\linewidth}\raggedright
\footnotesize
\textsuperscript{a)}Computed with $p=1$, the order of the calibration polynomial in Eq.~\eqref{eq:cal}.\\
\textsuperscript{b)}Number of wavelengths at which the platform is compared with the reference instrument within the common 400--700~nm range.\\
\textsuperscript{c)}Total number of calibration coefficients fitted for the platform.\\
\textsuperscript{d)}Standard deviation of the one hundred consecutive measurements, averaged over wavelength and over the three validation dyes.\\
\textsuperscript{e)}Coefficient of variation corresponding to SD, computed at points whose mean absorbance exceeds 0.05~AU.
\end{minipage}
\end{table}

\begin{figure*}[!htbp]
\centering
\includegraphics[width=\linewidth, draft=false]{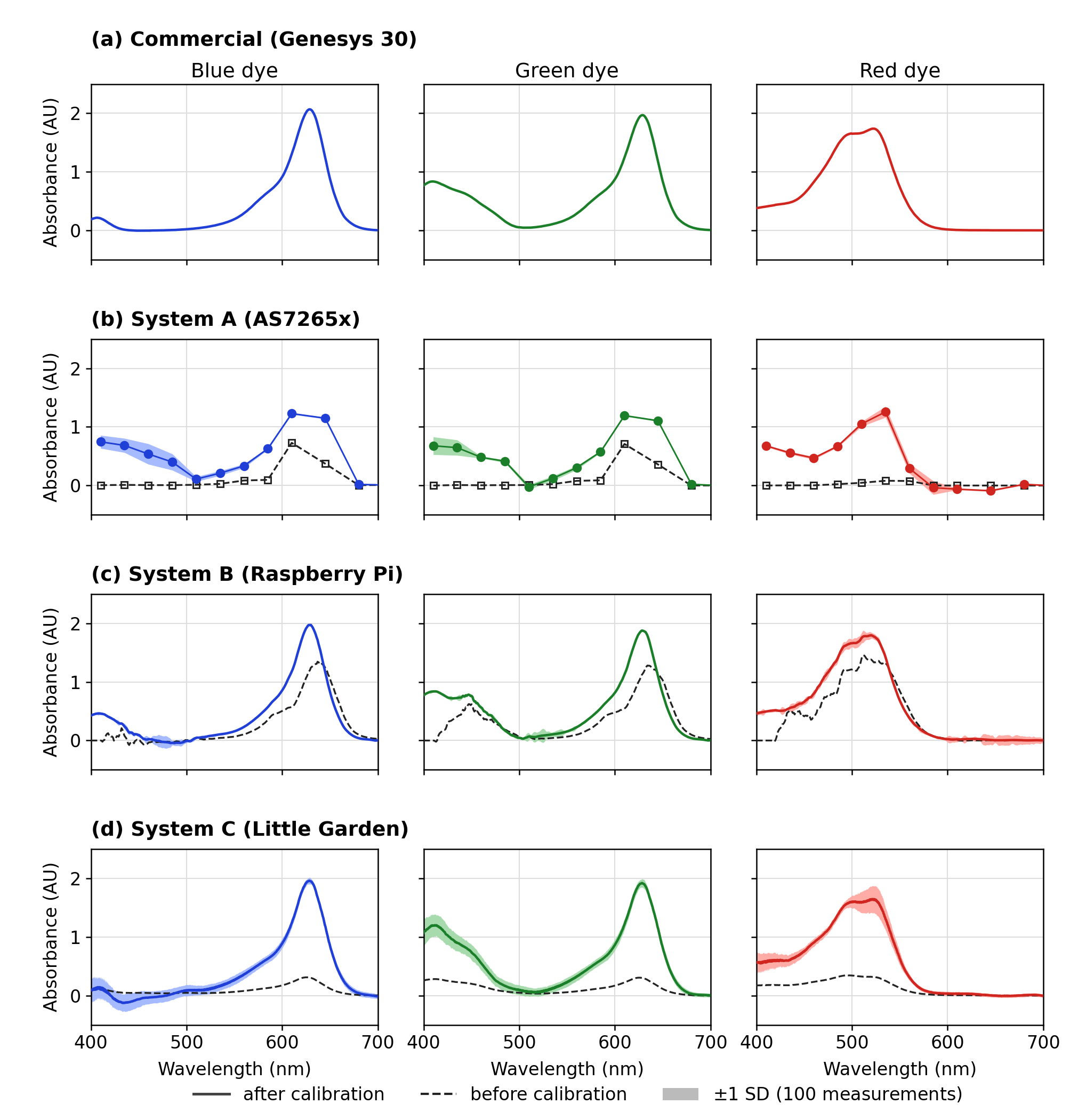}
\caption{Absorbance spectra of the blue, green and red validation dyes. (a) Commercial Thermo Scientific Genesys 30 reference spectrophotometer; (b) System A (AS7265x), (c) System B (Raspberry Pi) and (d) System C (Little Garden), each shown before the absorbance calibration (black dashed line) and after it (coloured line). For System A, symbols mark the 11 AS7265x channels that fall within the 400--700~nm range; the lines joining them are guides to the eye only. Shaded areas represent $\pm 1$ standard deviation obtained from 100 repeated measurements (not the standard error of the mean).}
\label{fig:comparison}
\end{figure*}

Figure~\ref{fig:comparison} compares the absorbance spectra obtained with the three low-cost spectrophotometric platforms against those measured using the commercial Thermo Scientific Genesys 30 spectrophotometer. Blue, green, and red solutions are presented as representative examples because they exhibit clearly distinct absorption bands within the visible range. Each platform is shown both before and after the absorbance calibration, so that the effect of the calibration can be judged directly: the uncalibrated spectra already reproduce the position of the absorption bands, but they underestimate their amplitude, most severely for Systems A and C, whose raw response spans a considerably narrower absorbance range than the reference instrument. For each sample, the shaded regions represent one standard deviation calculated from the one hundred consecutive measurements, and not the standard error of the mean, providing a visual estimate of each system's repeatability. The corresponding coefficients of variation are reported in Table~\ref{tab:metrics}.

\FloatBarrier

The three systems were able to correctly identify the main absorption bands of the analyzed samples. However, significant differences were observed with respect to spectral smoothness, reconstruction fidelity, and noise level.

\FloatBarrier

System A adequately reproduced the general spectral behavior of the reference instrument, despite using only eighteen discrete spectral channels. The main absorbance peaks were correctly identified in all three evaluated solutions, although the low sampling density limited the reconstruction of fine spectral details and smooth transitions between wavelengths. Consequently, it exhibited the lowest average determination coefficient with respect to the commercial spectrophotometer ($R^{2}=0.805$, $R^{2}_{adj}=0.783$; Table~\ref{tab:metrics}). This result reflects primarily the discrete multispectral sampling strategy of the AS7265x sensor rather than an intrinsic limitation of the instrument. Because only eighteen spectral channels are available, point-by-point agreement with a continuous reference spectrum is inherently constrained.

System B produced spectra with a considerably higher level of detail, thanks to an architecture based on imaging spectroscopy and continuous spectral acquisition. The reconstructed curves faithfully followed the shape of commercial spectra, although a moderate level of noise was observed, particularly in the blue region and, to some extent, in the green. The average coefficient of determination obtained ($R^{2}=0.983$, $R^{2}_{adj}=0.983$) is the highest of the three platforms, and reflects the balance between higher spectral resolution and the increased sensitivity of this type of system to optical alignment, detector noise, and calibration uncertainties. Its repeatability was intermediate between those of the other two platforms, with an average coefficient of variation of 4.0\% over the three validation dyes (Table~\ref{tab:metrics}).

System C reached a level of agreement comparable to that of System B, with an average coefficient of determination of $R^{2}=0.975$ ($R^{2}_{adj}=0.975$), and it did so with by far the lowest dispersion between repeated measurements: an average coefficient of variation of 0.3\%, an order of magnitude below the other two platforms. It also required the most compact calibration of the three, 18 coefficients against 38 for System B. Despite utilizing a simpler optical configuration—based on a diffraction grating derived from a DVD and a modified webcam—the reconstructed spectra closely matched those of the reference instrument. This result indicates that overall system performance depends not only on spectral resolution but also on the stability of the optical configuration and the effectiveness of the spectral reconstruction process.

\begin{figure*}[!htbp]
\centering
\includegraphics[width=\linewidth, draft=false]{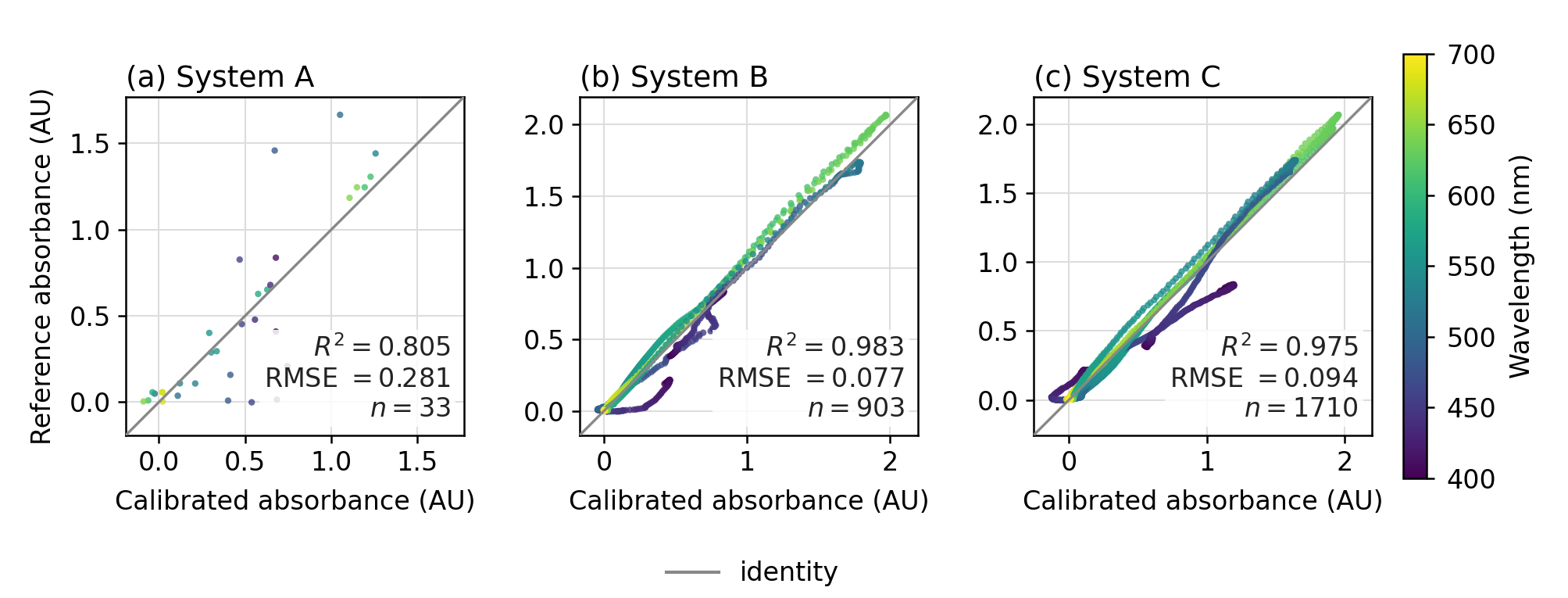}
\caption{Calibrated absorbance of each platform against the absorbance of the reference spectrophotometer, for the three validation dyes over the 400--700~nm range. Each point is one wavelength of one dye, coloured by that wavelength; the grey line is the identity. (a) System A (AS7265x), (b) System B (Raspberry Pi) and (c) System C (Little Garden). Each panel quotes the coefficient of determination reported in Table~\ref{tab:metrics}, the root-mean-square difference with respect to the reference over all the points of the panel, and the number of comparison points.}
\label{fig:map}
\end{figure*}

Table~\ref{tab:metrics} reduces the agreement of each platform to a single number. Figure~\ref{fig:map} shows the comparison that those numbers summarise: the calibrated absorbance of each platform against the absorbance measured by the reference instrument, for the three validation dyes, with every point coloured by the wavelength at which it was acquired. A platform in perfect agreement with the reference would place all of its points on the identity line.

Systems B and C collapse onto the identity line over most of the range, with root-mean-square differences from the reference of 0.077 and 0.094 respectively, whereas System~A remains a scattered cloud at 0.281.

The departures from the identity line are not distributed uniformly over the spectrum. In the three platforms they concentrate below 480~nm, where the mean absolute difference with respect to the reference is 2.1, 2.3 and 3.4 times larger than above that wavelength for Systems A, B and C respectively. That the same effect appears in three different architectures at once points to the halogen source, whose emission falls steeply towards the blue, rather than to any individual detector.

Table~\ref{tab:metrics} collects, for each platform, the coefficient of determination, its adjusted counterpart, the number of comparison points, and the dispersion between repeated measurements. The difference between $R^{2}$ and $R^{2}_{adj}$ is negligible for the two continuous platforms, where several hundred comparison points are available, but it is substantial for System A, whose eleven points within the common spectral range make the unadjusted value noticeably optimistic. The two orderings do not coincide: System B agrees best with the reference instrument, but System C is by a wide margin the most repeatable, and System A is the last on both counts.

These results demonstrate that spectral resolution, while important, is not in itself a predictor of a spectrophotometer's overall performance. The final quality of the measurements results from the interaction between the optical design, detector characteristics, mechanical stability, and the implemented processing algorithms.

\subsection{Practical Implementation and Cost Analysis}\label{sec4.1}

Analytical performance is just one of the criteria to consider when selecting a low-cost spectrophotometer. Aspects such as implementation complexity, portability, ease of calibration, and total cost are equally decisive in real-world applications.

Table~\ref{tab:cost} summarizes the estimated implementation cost of the three evaluated platforms.

\begin{table}[!htbp]
\caption{Estimated cost breakdown of the three systems}\label{tab:cost}
\centering
\footnotesize

\begin{tabularx}{\textwidth}{@{}>{\raggedright\arraybackslash}p{4.2cm} X X X@{}}
\toprule
Component & System A (\euro) & System B (\euro) & System C (\euro) \\
\midrule
Spectral sensor / camera & 65.00 & 15.00 & - \\
Microcontroller / SBC & 8.00 & 65.00 & - \\
Optical lens(es) & - & 15.00 & - \\
Diffraction grating & - & 100.00 & - \\
3D enclosure + accessories & 10.00 & 10.00 & - \\
Wiring, PCB, power supply & 5.00 & 10.00 & - \\
\midrule
Total estimated cost & 88.00 & 215.00 & 90.00 \\
\bottomrule
\end{tabularx}
\end{table}

System A presented the lowest implementation cost (\euro{} 88), as well as the simplest optical configuration and the greatest portability. These characteristics make it a particularly suitable option for field monitoring, environmental tracking programs, and educational applications where robustness and ease of use are prioritized over maximum spectral resolution.

System B required the largest investment (\euro{} 215), primarily due to the use of a commercial diffraction grating and a Raspberry Pi platform. Although it offers superior spectral resolution and great flexibility for experimental development, these advantages entail more demanding calibration procedures, more careful optical alignment, and higher implementation costs.

System C represented an intermediate solution in terms of implementation effort, while maintaining a total cost comparable to that of System A (\euro{} 90). Its high level of agreement with the commercial spectrophotometer demonstrates that relatively simple hardware can achieve high analytical performance when supported by an appropriate optical design and signal-processing strategy.

Overall, the results strongly suggest that no single low-cost spectrophotometric architecture can be considered universally superior. Each platform optimizes different design objectives; therefore, the most suitable choice depends on the specific requirements of the application. Multispectral sensors prioritize simplicity, portability, and low energy consumption, whereas image-based architectures provide a greater amount of spectral information at the cost of more complex implementation. Consequently, the choice of a low-cost spectrophotometer should be based on application needs rather than solely on a single performance indicator.

\FloatBarrier
\section{Conclusion}\label{sec5}

Here we have presented a comparative evaluation of three representative open-source, low-cost spectrophotometric platforms encompassing the two dominant architectures identified in the literature: discrete multispectral sensing and continuous-spectrum imaging spectroscopy. Unlike studies focused on the development of a single instrument, this work provides a systematic assessment of the strengths and limitations of each architecture using a common experimental protocol and a commercial spectrophotometer as the reference instrument.

The results demonstrate that no evaluated platform can be regarded as universally superior. The AS7265x-based system A stood out for its simplicity, low cost, low power consumption, and ease of implementation, making it well suited for field measurements, routine monitoring, and educational applications. Conversely, the imaging-based architectures (systems B and C) achieved greater spectral fidelity with respect to the commercial instrument. In particular, the Raspberry Pi platform (system B) obtained the highest average coefficient of determination ($R^{2}=0.983$), while the Little Garden platform (system C) reached a comparable figure ($R^{2}=0.975$) with the best repeatability by an order of magnitude and the most compact calibration of the three, demonstrating that an appropriate calibration strategy can partially compensate for the limitations of a simple optical configuration. Each platform is calibrated with between 18 and 38 fitted coefficients, which makes the procedure both reproducible and portable to other units of the same design.

An important outcome of this study is that spectral resolution alone is not a reliable predictor of spectrophotometer performance. Measurement quality results from the combined influence of optical design, calibration methodology, mechanical stability, signal processing, and overall system architecture. Therefore, the selection of a low-cost spectrophotometric platform should be driven by application requirements rather than by a single performance metric. It is hoped that the comparison methodology presented here will serve as a reference for the design and selection of future low-cost spectrophotometers intended for analytical, environmental, and educational applications.

\begin{acknowledgments}
We thank Axel Cornu for valuable advice. This publication presents the results of research obtained within the framework of the project ``In situ monitoring of water quality in Cuban bays: creating and promoting the use of a scientific toolbox'' supported by ARES with funding from the Belgian Development Cooperation. A.C. and S.C.N. acknowledge funding from the European Union's Horizon Europe research and innovation program under grant agreement No.~101181363 (BioDiMoBot).
\end{acknowledgments}

\section*{Author Declarations}

\subsection*{Conflict of Interest}
The authors have no conflicts to disclose.

\subsection*{Author Contributions}
\textbf{Alex M. Rivera-Rivera}: Conceptualization; Data curation; Formal analysis; Investigation; Methodology; Project administration; Resources; Software; Validation; Visualization; Writing -- original draft; Writing -- review \& editing.
\textbf{Daniela Martinez-Ortiz}: Conceptualization; Data curation; Formal analysis; Investigation; Methodology; Project administration; Resources; Software; Validation; Visualization; Writing -- original draft; Writing -- review \& editing.
\textbf{Fran\c{c}ois Quitin}: Investigation; Methodology; Writing -- review \& editing.
\textbf{Denis Terwagne}: Investigation; Methodology; Writing -- review \& editing.
\textbf{Stamatios C. Nicolis}: Conceptualization; Formal analysis; Investigation; Methodology; Visualization; Writing -- review \& editing.
\textbf{Ernesto Altshuler}: Conceptualization; Investigation; Methodology; Project administration; Supervision; Writing -- review \& editing.
\textbf{Alexandre Campo}: Conceptualization; Formal analysis; Investigation; Methodology; Project administration; Supervision; Visualization; Writing -- review \& editing.

\section*{Data Availability}
The data that support the findings of this study are available from the corresponding author upon reasonable request. The software developed in this work is openly available: the firmware and control software of System A at \url{https://gitlab.com/fablab-ulb/projects/cuban-water-lab/spectrophotometer-software}, and the acquisition and processing software of System B at \url{https://gitlab.com/fablab-ulb/projects/cuban-water-lab/pyspectrometer2}. System C is a commercial instrument operated with the manufacturer's software, so no repository is associated with it.

\bibliography{references}

\end{document}